# Diverse fundamental properties in stage-n graphite alkali-intercalation compounds: anode materials of Li$^+$-based batteries

Wei-bang Li[1], Ngoc Thanh Thuy Tran[2], Shih-yang Lin[3], Ming-Fa Lin[1,2]
[1]Department of Physics, National National Cheng Kung University, Tainan, Taiwan
[2]Hierarchical Green-Energy Materials (Hi-GEM) Research Center, National Cheng Kung University, Tainan 70101, Taiwan
[3]Department of Physics, National Chung Cheng University, Chiayi, Taiwan

## Abstract

The diversified essential properties of the stage-n graphite alkali-intercalation compounds are thoroughly explored by the first-principles calculations. According to their main features, the lithium and non-lithium materials might be quite different from each other in stacking configurations, the intercalated alkali-atom concentrations, the free conduction electron densities, and the atom-dominated & (carbon, alkali)-co-dominated energy bands. The close relations between the alkali-doped metallic behaviors and the geometric symmetries will be clarified through the interlayer atomic interactions, in which the significant alkali-carbon chemical bondings are fully examined from the atom- and orbital-decomposed van Hove singularities. The blue shift of the Fermi level, the n-type doping, is clearly identified from the low-energy features of the density of states. This study is able to provide the partial information about anode of Li$^+$-based battery. There are certain important differences between $AC_6/AC_8$ and $Li_8Si_4O_{12}$.

## 1. Introduction

A bulk graphite[1], which could be regarded as the layered graphene systems [2], has attracted a lot of theoretical and experimental researches in the basic science [3-5], engineering[6] and application[7]. Each graphene layer is a pure carbon honeycomb lattice[8]; furthermore, the graphitic are attracted together through the weak, but significant van der Waals interactions[9]. There exist the very strong intralayer $\sigma$ bondings of C-[2s, 2px, 2py] orbitals and the important interlayer C-2pz orbital hybridizations[10], in which the former determine the most outstanding mechanical material , and the latter dominate the low-lying energy bands and thus the essential physical properties . Very interesting, the 3D graphite systems might present the AA[11], AB[12], ABC[13] and turbostratic stackings[14]. All of them belong to semimetals under the interlayer hopping integrals of C-2pz orbitals. In general, such condensed-matter systems become the n- or p-type metals, depending on the kinds of intercalated atoms or molecules. For example, the intercalation of alkali atoms[15] and $FeCl_3$ into graphite, respectively, create many free conduction electrons and valence holes, as observed in a pure metal.

It is well known that graphite could serve as the best anode materials[16] in the commercialized $Li^+$ based batteries[17], mainly owing to the lowest cost, the most stable structure for intercalation, and the outstanding ion transport under the charging and discharging processes. When the $Li^+$ ions are released from the cathode, they will transport through the electrolyte, and then intercalate into the graphitic system. Most important, the flexible interlayer spacings between graphene layers are capable of providing the sufficient positions for the various intercalant concentrations[18]. Any intermediate states & meta-stable configurations, which are created during the ion/atom intercalation, could survive through the very strong $\sigma$ bondings in graphitic sheets. As a result, graphite is rather suitable for studying the structural transformation in the chemical reactions. For example, the close relations between graphene and intercalant layers in lattice symmetries are expected to present the dramatic transformation before and after the intercalation/de-intercalation processes.

Up to now, there are a lot of theoretical and experimental studies on the fundamental properties in graphite-related systems. The former covers the phenomenological models and number simulation methods. For example, the tight-binding model, the random-phase approximation, and the Kubo formula are, respectively, utilized to investigate their electronic properties & magnetic quantization behaviors[19-22], Coulomb excitations[23-27] & impurity screenings, and optical absorption spectra[28]. Furthermore, the first-principles calculations[29] are available in understanding the optimal stacking configurations, intercalant lattices, $\pi$, $\sigma$ & intercalation-induced energy bands, free conduction electron/valence hole densities, and density of states. However, certain critical physical quantities and pictures are absent in the previous studies, such as, the atom-dominated band structures, the spatial charge densities between intercalant and graphene layers, the interlayer orbital hybridization of intercalant and carbon atom, the atom- & orbital-projected van Hove singularities, and the spin-dependent state degeneracy/magnetic moment/density distribution.

The main focuses of this book chapter are the main features of geometric and electronic properties in stage-n graphite alkali-intercalation compounds using the first-principles method[29]. Very interesting, the dependences on the kinds of intercalants and their concentrations are thoroughly explored through the delicate calculations and analyses. The stacking configurations of graphene sheets, the lattice symmetries of intercalant layers, the dominances of valence and conduction states by the carbon and/or alkali atoms, the interlayer charge density variations after the alkali intercalations, and atom- & orbital-decomposed density of states. Whether the spin-induced Coulomb on-site interactions or the spin-orbital couplings play critical roles will be examined in detail. Most important, the orbital-hybridizations in chemical bonds and the atom-induced spin configurations are determined from the above-mentioned results. How to simulate VASP band structures by the tight-binding model is worthy of detailed discussions, since the interlayer hopping integrals are not very complicated under the small primitive unit cells of layered graphite compounds. The predicted interlayer distance/intercalant lattice, band overlap, and unusual van Hove singularities, could be detected using the high-resolution X-ray diffraction/low-energy electron diffraction[30], angle-resolved photoemission spectroscopy[31], and scanning tunneling spectroscopy[32], respectively.

## 2. Theoretical calculations

Both numerical simulations and phenomenological models are available in fully exploring the diversified physical/chemical phenomena. For example, the former and the latter, respectively, cover the first-principles calculations[29]/molecular dynamics[33]/quantum Monte Carlo[34] and the generalized tight-binding model[35]/Kubo formula[36]/random-phase approximation[37]. Apparently, only the first method is able to investigate the lattice symmetries of the graphite-related systems with/without the chemical modifications. As for pure graphites and graphite alkali-intercalation compounds, the previous VASP results show that the stacking configurations and the intercalant lattice symmetries & concentrations play important roles in determining the semi-metallic or metallic behaviors. However, the concise pictures, which are very useful in fully comprehending the fundamental properties, are required in the further studies, as done for 2D layered graphenes[38,39] and 1D graphene nanoribbons[40-42].

It is well known that the Vienna ab initio simulation package is developed within the density functional theory[43]. The up-to-date VASP calculations can provide the sufficient information on the lattice symmetries, valence & conduction bands, charge density distributions/variations after chemical bondings, atom- & orbital-decomposed density of states[44], being accompanied with/independent of the spin-related properties. Very interesting, a lot of electron quasiparticles experience two very complex interactions, when their wave packets propagate in any periodical material. Beyond the classical electrodynamics, the many-body electron-electron Coulomb interactions could be classified into the exchange and correlation energies according to the first-order and the other corrections, respectively. The Perdew-Burke-Ernzerh formula, which depends on the local electron density, is utilized to deal with many-particle Coulomb effects. As for the frequent electron-crystal scatterings, they are characterized by the projector-augmented wave pseudopotentials. The electron Bloch wave functions are solved from the linear superposition of plane waves, with the maximum kinetic energy of 500 eV. The current study on stage-n graphite compounds shows that the first Brillouin zone is sampled by $9 \times 9 \times 9$ and $100 \times 100 \times 100$ k-point meshes within the Monkhorst-Pack scheme, respectively, for the optimal geometry and band structure. Moreover, the convergence condition of the ground state energy is set to be $\sim 10^{-5}$ eV between two consecutive evaluation steps, where the maximum Hellmann-Feynman force for each ion is below 0.01 eV/Å during the atom relaxations.

The self-consistent calculations, which are closely related to the local carrier density[45], account for the VASP simulations through the Kohn-Sham equations. According to the initial charge distribution, one is able to evaluate the classical & many-particle exchange-correlation Coulomb interactions, and the electron-ion scattering potentials. After solving the second-order differential equations, the intermediate electron density and ground state energy are obtained and then are utilized to examine the convergence condition on the Feynman force. The significant physical quantities, the lattice symmetries/the various bond lengths in a unit cell, the atom-dominated valence & conduction bands, the spatial charge densities/their variations after the creation of chemical bonds, the atom- & orbital-projected van Hove singularities, the atom-induced spin distributions, the spin-split or degenerate states across the chemical potential, and the vanishing or non-vanishing magnetic moments, are available in making a final decision about the critical multi-/single-orbital hybridizations of chemical bonds and the atom-induced spin configurations in the stage-n graphite alkali-intercalation compounds. The above-mentioned theoretical framework, which is built from the first-principles results, presents the successful exploration on the diversified essential properties of 2D layered graphenes[38,39], 1D graphene nanoribbons[40-42], and 2D silicene-based systems. Now, the developed viewpoints are suitable for the anode, cathode and electrolyte materials of $Li^+$-based batteries[46]. For example, a detailed comparison between graphite and $Li_4Ti_5O_{12}$ is interesting and meaningful in term of well-known anode compounds.

The direct linking between numerical simulations and phenomenological models is a very interesting topic. According to the up-to-date results of the low-energy band structures, the combination of the VASP calculations and the tight-binding model is rather successful in exploring layered graphene/graphitic systems. For example, both methods can account for the linear and gapless Dirac-cone band structure, two pairs of linear/parabolic valence and conduction bands, three vertical Dirac cones/two pairs of parabolic bands plus one separated Dirac cone/one pair of partially flat, sombrero-shape and linear energy dispersions/oscillatory, sombrero-shape and parabolic bands, respectively, for the monolayer, bilayer AA-/AB-stacked, and trilayer AAA-/ABA-/ABC-/AAB-stacked graphene systems. Their results are consistent with each other, since the important interlayer atomic interactions are purely due to the carbon-2pz orbitals within the specific unit cells. That is to say, the single-orbital Hamiltonians, with the concise interlayer & intralayer hopping integrals, could be expressed in the analytic forms. Furthermore, all the intrinsic interactions, being modified by the external fields, are further included in the generalized tight-binding model. This theoretical framework has been successfully in creating and explaining the rich and unique magnetic quantization phenomena of few-layer group-IV systems, e.g., the unusual quantized Landau levels in graphene, silicene[47], germanene[48] and tinene[49].

## 3. Unique stacking configurations and intercalant distributions

The normal stacking configurations, which are formed by the layered graphene systems, cover AAA, ABA, ABC and AAB. For example, according the VASP calculations on the trilayer graphenes, the ground state energy/the interlayer distance is lowest and highest/shortest/longest for the ABA and AAA stackings. The similar results are revealed in the AA-, AB- and ABC-stacked graphites. Each honeycomb structure has two equivalent sublattices with a planar geometry, indicating the orthogonal relation between one 2pz and three [2s, 2px, 2py] orbitals. The $sp^2$-bonding-based graphene systems maintain the flat sheets even under any interlayer couplings. The significant van der Waals interactions, the interlayer single-orbital hybridizations of 2pz-2pz, account for the stacking configurations. Also noticed that the 2D silicene[50]/germanene[51]/tinene[52] systems possess the buckled structures, since the $\pi$ and $\sigma$ bondings are non-orthogonal to each other. Furthermore, the $sp^3$ bonding might play an important role in the interlayer atomic interactions. Apparently, graphenes are quite different from silicenes/germanenes/tinenes in various essential properties.

The interlayer spacings of graphitic layers are able to provide the very suitable chemical environment for the alkali intercalations and de-intercalations, i.e., they are very useful during the charging and discharging processes. The chemical modifications hardly affect the planar honeycomb lattices and thus do not change the orthogonal $\pi$ and $\sigma$ bondings. Very interesting, the optimal geometric properties are rather sensitive to the (x, y)-plane distribution and concentration of alkali atoms, e.g., the interlayer distances of neighboring planes and an enlarged primitive unit cell. According to the highest concentration, the graphite alkali-intercalation materials could be classified into two categories: lithium and non-lithium ones. There exist $LiC_{6n}$ or $AC_{8n}$ for the stage-n materials [Figs. 7.1(a)-7.1(d)], as examined from the experimental measurements and the theoretical predictions. The specific distance between carbon and intercalant layer is shortest/longest for the Li/Cs [Table 7.1], only directly reflecting the effective atomic radius. Although all the alkali atoms present the hollow-site optimal positions, their distribution symmetries are dominated by Li or non-Li. The layered

$LiC_{6n}$ and $AC_{8n}$ systems, respectively, possess the three- and four-times enhancement in the unit cell; therefore, their reduced first Brillouin zones exhibit the different high-symmetry points [Fig. 7.3(a)]. This is expected to have a strong effect on the initial π-electronic states of the Dirac-cone structure. More chemical bonds, which arise from the alkali atoms, are produced in graphite intercalation compounds, compared with pristine systems. Such bondings need to be taken into account for the diversified essential properties. In addition to the intercalant distribution, the stacking configuration of two neighboring graphitic sheets also affects the ground state energy. The current study clearly shows that only the stage-1 $LiC_6$ presents the AA stacking, and other compounds possess the AB stacking. That is to say, all the graphite alkali-intercalation materials exhibit the AB-stacked configurations except for the former system.

The high-resolution experimental measurements of X-ray scatterings and low-energy electron diffractions[30] are able to examine the intercalant distribution, interlayer distance, and stacking configuration of graphite alkali-intercalation compounds; furthermore, those of scanning tunneling microscopy and tunneling electron microscopy are very suitable for the identifications of optimal geometric structures in few-layer graphene system[37]. The previous experiments have shown the AA/AB stacking of the stage-1 $LiC_6$/$AC_8$ material, while the predicted AB stacking in stage-2-4 compounds [$LiC_{12}$, $LiC_{18}$ and $LiC_{24}$] require the further experimental verifications.

**4. Metallic and semi-metallic behaviors**

Without alkali atom intercalations, a pristine graphite exhibits the unique semi-metallic behavior through the weak, but significant van der Waals interactions. It is well known that a monolayer graphitic sheet has the orthogonal π and σ bondings, in which the initial electronic states, respectively, come to exist at the Fermi level [$E_F=0$] and E~-3.10 eV. Obviously, the low-energy physical properties are mainly determined by the intralayer atomic interactions. Such hopping integrals, being accompanied with the uniform three nearest neighbors are responsible for a gapless and isotropic Dirac cone initiated from the K/K' valleys. The density of states is vanishing at $E_F$ because of the specific Dirac points. A semiconducting zero-gap graphene is dramatically changed into a semi-metallic system for any few-layer stackings, or infinite-layer graphites. For example, there exist the unusual overlaps of valence and conduction bands along the KH path [or the Γ path under the reduced Brillouin zone] in AA-, AB- and ABC-stacked graphites under the single-orbital interlayer atomic interactions. According to the VASP and tight-binding model calculations, the simple hexagonal/rhombohedral graphite [Fig. 7.3(a)], with the highest/lowest stacking symmetry, presents the largest/lowest free carrier density of conduction electrons and valence holes. In general, the π valence bands, which are purely related the C-2pz, show the initial states at the stable K valley, the saddle-point structure near the M point [$E^v$~-3eV], and its termination in the Γ valley [the band width more than 7 eV.

Very interesting, the band structures of lithium- and non-lithium graphite intercalation compounds sharply contrast with each other, such as, those of stage1 $LiC_6$ and $XC_8$ compounds in Figs. 7.3(d) and 7.3(e)-7.3(h), respectively. For the former, the Fermi level is transferred from the middle of Dirac-cone structures into the conduction ones near the Γ valley after the intercalation of lithium atoms. That is to say, $E_F$ presents a blue shift. Apparently, the free carriers are due to the outmost 2s orbitals of lithium atoms. The occupied and unoccupied states are highly asymmetric to each other about the Fermi level. The initial π/π* valence/conduction states are mainly determined by the stacking configurations in the chemical intercalations [the distribution symmetry of Li-intercalants in Fig. 7.2(a)], or the corresponding relation between the original and reduced first Brillouin zone. There exists an observable energy spacing of $E^c$~0.31-0.63 eV along the Γ-A path in the modified valence and conduction Dirac-cone structures. The creation of discontinuous states in honeycomb lattices might arise from the different ionization energies of Li-2s and C-2pz orbitals [the distinct on-site Coulomb potential energies. The whole π valence-band width of $LiC_6$, being created by the C-2pz orbitals, could be identified to about 7.32 eV/7.40 eV from the electronic energy spectrum along the Γ-M-K-Γ/A-H-L-A [or Γ-K-M-Γ/A-L-H-A] paths. This result clearly illustrates the well-behaved π bondings in graphite alkali-intercalation compounds. Another σ bondings behave so, as indicated from the initial states at the Γ point of $E^v$~-4.20 eV. Their orthogonal relation remains the same after the chemical modification, being very useful in establishing the tight-binding model for graphite alkali-intercalation compounds.

The main features of band structure, as clearly indicated in Figs. 7.3(e)-7.3(h), are dramatically changed under the other alkali-atom intercalations. Both stage-1 $AC_8$ and $LiC_6$ have the totally different distribution configurations [Figs. 7.2(a) and 7.2(b)], and so do the reduced first Brillouin zones [Figs. 7.3(a)]. For the former, the Dirac-cone structure of the π and π* bands are initiated from the stable K/K' valleys, but not the Γ ones. The energy spacing of separated Dirac points is small or almost vanishing; furthermore, the Fermi level is situated above the conduction point about ~1.35-1.50 eV. In addition to $E_F$, whether the second conduction energy subband is partially occupied depends on the kinds of alkali atoms, such as, the alkali-induced free electrons in the first and second subbands for $KC_8$/$RbC_8$/$CsC_8$ [Figs. 7.3(f)/7.3(g)/7.3(h)]. The low-energy bands belong to the single states, without the split double degeneracy [Fig. 7.3(d) for $LiC_6$]. The whole π-band energy spectra are identified from the K-Γ-K-M-Γ and H-A-H-L-A paths, leading to the width of ~ 7.31-7.41 eV. Very interesting, one pair of π-valence subbands come to exist near $E^v$~-4.0 eV, being accompanied with the initial pair of the σ valence bands. This further illustrates the zone-folding effects on band structure and the well separation of π and σ bondings.

Electronic energy spectra strongly depend on the n stage of graphite intercalation compound, as clearly illustrated in Figs. 7.4(a)-7.4(h). The stage-2 $LiC_{12}$ and $AC_{16}$ have the different first Brillouin zones on the $(k_x, k_y)$ plane, and their $k_z$-ranges are associated with the distances between two intercalant planes [Fig. 7.1]. Compared with stage-1 band structures [Figs. 7.3(d)-7.3(h)], the number of Dirac cones becomes double in the stage-2 cases, in which the further modifications cover the reduced blue shift of the Fermi level, the enhanced anisotropy, the induced energy spacings between valence and conduction Dirac points, and the diversified energy relations near the band-edge states. For example, four/two valence and conduction pairs come to exist from the Γ/K valley [or the A/H valley] for $LiC_{12}$/$AC_{16}$ [Fig. 7.4(d) and Figs. 7.4(e)-7.4(h)]. Furthermore, the whole π-band widths could be roughly estimated from the ΓKMΓ /KMKΓ path [or the ALHA/HLHA path]. The low-energy essential properties are dominated by the π bondings of C-2pz orbitals. The above-mentioned obvious changes directly reflect the great enhancement of the interlayer atomic interactions due to the C-C bonds in two neighboring graphitic sheets/graphene-intercalant layers.

Very interesting, all the stage-2 compounds exhibit a pair of $\sigma$ bands at the Γ and A valleys, and the energy dispersions along Γ-A are negligible. These results indicate the mutual orthogonality of the planar $\sigma$ and perpendicular π bondings. Such phenomenon is expected to survive in any stage-n graphite alkali-intercalation compounds.

The blue shifts about the Fermi level and the free conduction electrons declines quickly in the increase of the n number. The concentration of alkali atoms is greatly reduced, and do the charge transfer from their outmost s-orbitals to carbon 2pz-ones. Such results are clearly revealed in the stage-3 systems. Figs. 7.5(a)-7.5(f) show the slight modifications of the Dirac-cone structures and the intersecting of the Fermi level with most of conduction bands. $LiC_{18}$ [Fig. 7.5(b)] and $AC_{24}$ [Figs. 7.5(c)-7.5(f)], respectively, possess six and three pairs of linear valence and conduction bands near the Γ/A and K/H valleys. Apparently, there exist the significant changes in the observable energy spacing between valence and conduction Dirac points, the anisotropic Fermi velocities, and the distinct Fermi momenta slopes. The blue shifts of $E_F$ are estimated to be 0.45 eV, 0.58, 0.55, 0.53, and 0.5 eV's. Very interesting, the electronic energy spectra become more complicated under the stronger zone-folding effects. The main features of electronic properties are closely related to the weak, but important Li-C bonds, e.g., the minor contributions of alkali atoms on each electronic states. That is, the diversified n-type dopings are created by the critical Li-C and C-C bnondings.

The conduction electron density, which is created by the alkali-atom doping, is worthy of a closer examination. It is mainly determined by the covered volume [area/length] in the wave-vector space between the lowest conduction state and the Fermi level for 3D graphite alkali-intercalation compounds[53] [2D alkali-adsorbed graphhenes/1D graphene nanoribbons]. In general, the VASP calculations are too heavy to obtain the delicate values because of the strong anisotropic energy dispersions [the non-linear Dirac-cone structure at higher energies in Figs. 7.4(d)-7.3(h)]. That a lot of 3D **k**-points are required for the accurate calculation of Free electron density is almost impossible. The similar phenomenon is revealed in the alkali-adatom adsorptions on graphene. However, 1D graphene nanoribbons are very suitable for the full exploration of charge transfer and conduction electron density, since the Fermi momenta are proportional to the Latter. After examining the critical mechanisms, namely the zigzag/armchair edges, different widths, five kinds of alkali atoms, and their various distributions, the free electron density is almost identical to the adatom concentration[54]. It can be deduced that any one alkali adatom contributes the outmost half-occupied s-orbital as a conduction carrier, or it behaves the full charge transfer of -e in the alkali-carbon bonds. Such conclusions might be suitable for 2D and 3D alkali-doped graphite compounds, being supported by the following result. According to the spatial charge distributions and their variations after the chemical modifications, the chemical bondings hardly depend on the distributions and concentrations of alkali adatoms.

Generally speaking, only the high-resolution ARPES is capable of detecting the wave-vector-dependent occupied energy spectra below the Fermi level. The up-to-date experimental measurements have successfully verified the diversified phenomena for the graphene-related $sp^2$-bonding systems, as clearly observed under the various dimensions, layer numbers, stacking configurations, substrates, and adatom/molecule chemisorptions. For example, there exist 1D parabolic dispersions with sufficiently wide energy spacings/band gaps in graphene nanoribbons, the linear Dirac cone for monolayer graphene, two parabolic valence bands in bilayer AB stacking, the linear and parabolic dispersions in twisted bilayer systems, one linear and two parabolic bands for tri-layer ABA stacking, the linear, partially flat and Sombrero-shaped dispersions in tri-layer ABC stacking, the substrate-created observable energy spacing between the separated valence and conduction bands in bilayer AB stacking/oscillatory bands in few-layer ABC stacking[2], the semimetal-semiconductor/semimetal-metal transitions after the adaom/molecule adsorptions graphene surface, and the bilayer- and monolayer-like energy dispersions, respectively, at the K and H corner points [Fig 7.3(a)] for the Bernal graphite. The ARPES measurements are available in examining the main features of occupied band structure in graphite alkali-intercalation compounds, namely, the blue shift of the Fermi level, the initial K or Γ valley for the π & π* bands, the separated or almost close Dirac-cone structures, their double or single degeneracy, the whole π band along the Γ-K-M-Γ A-L-H-A paths [or the K-Γ-K-M-Γ /H-A-H-L-A paths], the formation of $\sigma$ subbands at E~-4.0 eV from the Γ valley, and the sensitive dependences on the alkali-atom distributions and concentrations. The experimental verifications could provide the useful information about the zone-folding effects due to the alkali interaction, the single-orbital hybridizations in alkali-carbon bonds, and the almost full charge transfer [discussed in the previous paragraph].

Density of states at E [DOS(E)], in which is defined the number of states within a very small energy range of dE, is capable of fully understanding the valence and conduction energy spectra simultaneously. Generally, its special structures are created by the band-edge states with the vanishing group velocities. The critical points in the energy-wave-vector space cover the local extreme points [minima & maxima], the saddle points, and all the partially flat energy dispersions. Apparently, the van Hove singularities are greatly diversified under the different dimensions. Of course, the discrete energy levels in 0D quantum dots only present the delta-function-like peaks, as observed in the magnetically quantized Landau levels in emergent layered materials[2] [e.g., few-layer graphene systems], and any-dimensional dispersionless energy bands. The 1D parabolic and linear bands, respectively, the asymmetric divergent peaks in the square-root form and the plateau structures. As for the 2D condensed-matter systems, the concave-/convex-form dispersions], the Dirac-cone structure [the linear valence and conduction bands without energy spacing] , the saddle points, and the constant-energy loops, respectively, create the broadening shoulders, V-shapes, the symmetric peaks relate related to the logarithmical divergence, and the square-root divergent peaks. Moreover, the higher dimension in 3D materials make them only show symmetric/asymmetric peaks and the shoulder structures, mainly owing to the absence of quantum confinement. The latter come from the 3D parabolic energy dispersions.

The delicate calculations and analyses, which are conducted on the atom- and orbital-dependent density of states [Figs. 7.6(a)-7.6(l)], are capable of fully understanding the metallic behavior and the close relations among the chemical bondings. Although band structures, with many band-edge states, become very complicated under the alkali-atom intercalations, the main features of van Hove singularities are sufficiently clear for the identifications of diversified phenomena through a suitable broadening factor [e.g., energy width of ~0.10 eV. This further illustrates a critical role of zone folding in the fundamental properties. Very interesting, the low-energy DOSs in stage-1 and stage-2 graphite alkali-intercalation compounds present a prominent peak just at the Fermi level [$E_F = 0$], regardless of the kind of alkali atoms. Furthermore, there exists a valley structure, with a minimum value, at its left-hand neighbor. When such characteristic is combined with the similar ones at $E_F$] in the pristine AA- and AB-stacked graphite systems [Figs. 7.6(a) and 7.6(b)], the Fermi level is deduced to exhibit a blue shift. That is to say, $E_F$ is situated at the conduction energy subbands [$E_F$ roughly lies in the center of valence and conduction bands] roughly after [before] the alkali-atom intercalations. It should be noticed that the contributions due to the alkali atoms are weak, but rather important. The blue shifts of the stage-1 $LiC_6$, $NaC_8$, $KC8$, $RbC_8$, & $CsC_8$, and stage-2 $LiC_{12}$, $NaC_{16}$, $KC_{16}$, $RbC_{16}$ & $CsC_{16}$ are, respectively, estimated to be 1.90, 1.80, 1.70, 1.60, 1.50, 1.30, 1.20, 1.10, 1.00, 0.90 eV's.

The van Hove singularities in graphite intercalation compounds [Figs. 7.6(c)-7.6(l)], which survive in the specific energy ranges, are mainly determined by the carbon or alkali atoms and their orbitals. Most important, DOS in the critical energy range of -5.0 eV < E < 3.0 eV, being relatively easily examined from the experimental STS measurements, are dominated by the carbon-2pz orbitals [the pink curves]. Furthermore, the minor contributions related to the outmost s-orbital of alkali atoms [the green curves], especially at the conduction energy spectrum, play a critical role in determining the blue shift of the Fermi level. As for the C-[2px, 2py] orbitals [the red and light blue curves], their contributions are initiated from ~ E < -4.0 eV, while they are absent in the opposite energy range. The red shift of the initial $\sigma$ valence bands is about 1.0 eV, compared with those of the pristine simple hexagonal and Bernal graphites [Figs. 7.6(a) and 7/6(b)]. Only the $LiC_6$ and $LiC_{12}$ cases [Figs, 7.6(c) and 7.6(h)] exhibit the split contributions of 2px and 2py orbitals. This result directly reflects the anisotropic distribution configuration. More, the C-2s orbitals com to exist at the deeper energies of ~E<-5.0 eV. Apparently, the above-mentioned features indicate the well separation of C-2pz and C-[2s, 2px, 2py] orbital contributions, and thus the normal perpendicular orbital hybridizations of $\pi$ and $\sigma$ chemical bonds. That such bonding behavior is strongly linked with the significant interlayer 2pz-orbitals due to the carbon-alkali bonds can account for the featured electronic properties, e.g., the main features of band structures and DOSs in Figs. 7.3-7.7].

The clear identifications of stage-1 and stage-2 of graphite alkali-intercalation compounds could be achieved from the low-energy features of van Hove singularities, as indicated in Figs. 7.6(c)-7.6(g) and 7.6(h)-7.6(l), respectively. Compared with those of the Former, the blue shifts of the Fermi levels are relatively small, in which they are, respectively, ~1.20 eV, 1.15 eV, 1.10 eV, 1.06 eV and 1.03 eV for $LiC_{12}$, $NaC_{16}$, $KC_{16}$, $RbC_{16}$, and $CsC_{16}$. Most important, their densities of states at $E_F$ do not belong to the local maxima. Such result directly reflects whether the band-edge states of conduction bands are somewhat away from the Fermi level [Figs. 7.4(b)-7.4(f)]. This significant difference between stge-2 and stage-1 systems further illustrates the intersecting of $E_F$ and conduction bands, thus leading to the n-type doping cases after the alkali-atom intercalations. The experimental measurements could be utilized to verify the free conduction electrons due to the alkali-atom intercalations, e.g., the atom-dependent optical threshold absorption frequencies, and the doping-enhanced electrical conductivities.

The high-resolution STS measurements are the only method in examining the van Hove singularities due to the valence and conduction energy spectrum, especially for the semi-metallic, metallic or semiconducting behaviors near the Fermi level. Such examinations cover the form, energy, number, and intensity of special structures in density of states. Apparently, they cannot identify the wave-vector-dependent energy dispersions. The up-to-date measured results have successfully verified the diverse electronic properties in graphene-related systems with the dominating $sp^2$ bondings, e.g., 1D graphene nanoribbons, carbon nanotubes, 2D few-layer graphenes, adatom-adsorbed graphenes, and 3D graphite. The main features of graphene nanoribbons, the width- and edge-dependent energy gaps and the square-root-divergent asymmetric peaks of 1D parabolic dispersions, are confirmed from the precisely defined boundary structures. The similar strong peaks are displayed in seamless carbon nanotubes, in which these structures exhibit the chirality- and radius-dependent band gaps and energy spacings between two neighboring subbands. Even with the obvious curvature effects on a cylindrical surface [the misorientation of 2pz orbitals and the significant hybridization of carbon four orbitals], armchair nanotubes belong to 1D metals with a finite density of states at the Fermi level [a sufficiently high free carrier density]. A lot of STS measurements, which are made for few-layer and adatom-doped graphenes, clearly illustrate the low-lying characteristics of van Hove singularities: a V-shape energy dependence vanishing at the Dirac point in monolayer system [a zero-gap semiconductor], the asymmetry-induced peak structures in the logarithmic form for twisted bilayer graphenes, a gate-voltage-created band gap in bilayer AB stacking and tri-layer ABC stacking , a delta-function-like peak centered about the Fermi level related to surface states of the partially flat bands in tri-layer and penta-layer ABC stackings, a sharp dip structure at $E_F$ accompanied with a pair of asymmetric peaks in tri-layer AAB stacking [a narrow-gap semiconductor with the low-lying constant-energy loops], and a red shift of Dirac point arising from the n-type electron doping of Bi adatoms. The measured density of states in Bernal graphite is shown to be finite near $E_F$ characteristic of the 3D semi-metallic property and presents the splitting of $\pi$ and $\pi^*$ strong peaks at deeper/higher energies. It should be noticed that the well-defined $\pi$ and $\sigma$ band widths in graphene-related systems are worthy of the further STS examinations. In addition, there are no STS results for AA- and ABC-stacked graphites up to now, mainly owing to few content of natural graphite and the difficulties in sample growth. The experimental examinations on the current predictions, especially for the van Hove singularities near $E_F$ and the $\pi$ & $\sigma$ band widths, are very useful in providing the detailed information about the n-type dopings and the orbital hybridizations of chemical bonds.

The close relations between the VASP simulations and the tight-binding model/the effective-mass approximation in graphite-/graphene-related systems are discussed in detail. The previous theoretical studies show that the electronic properties of pure carbon honeycomb lattices are well characterized by the phenomenological model. According to the calculated results of two different methods, the rich and unique band structures are created by the stacking configuration, layer number, and dimensionality, such as, those due to monolayer graphene, bilayer AA/AB stackings[55], trilayer AAA/ABA/ABC/AAB[56-58], 3D AA-/AB-/ABC-stacked graphites, 1D achiral/chiral carbon nanotubes[59], and graphene nanoribbons. Their outstanding consistence is deduced to be closely related to the good separations of $\pi$ and $\sigma$ chemical bondings and the interlayer single-2pz-orbital hybridizations. Very interesting, the generalized tight-binding model, which is combined with the static & dynamic Kubo formulas[60] and the modified random-phase approximation, is capable of fully exploring the diversified magnetic quantization. As for the emergent 2D group-IV and group-V materials, these theoretical frameworks are successful in studying the unusual magneto-electronic properties, the various magneto-optical selection rules, the unique quantum Hall conductivities, and complex magneto-plasmon modes & inter-Landau-level excitations[61].

The low-energy physical properties of graphites and graphite alkali-intercalation compounds are dominated by the single-orbital interactions of C-2pz and A-s orbitals [the outmost one of alkali atom]. The AA-, AB- and ABC-stacked graphites[11-13] have two, four and six carbon atoms in a unit cell, so their Hamiltonians are built from the corresponding tight-binding functions. After the diagonalization of Hermitian matrices, the π- and π*-electronic energy spectra strongly depend on the stacking configurations, especially for the overlap of valence and conduction bands. The highest conduction electron/valence hole density is revealed in the first/third system. Apparently, the stacking-enriched interlayer hopping integrals of C-2pz orbitals are responsible for the diversified semi-metallic properties. Most important, the above-mentioned phenomenological method could cover the alkali intercalation effects, the enlarged unit cells and the stacking configurations. That is to say, the modified Hamiltonians of graphite alkali-intercalation compounds include the ionization energy difference between C-2pz and Li-2s orbitals, the intralayer nearest-neighboring interactions of 2pz-2pz/2s-2s orbital in C-C/Li-Li bonds, and the interlayer 2pz-2s/2pz-2pz orbital hybridizations. Specifically, the third ones might be complicated and very sensitive to the change in the intercalant concentration; therefore, they play a critical role in creating the diversified metallic behaviors. It should be noticed that the theoretical models, with the zone-folding effects, are absent up to now.

In addition to geometric and electronic properties, free conduction electrons in graphite intercalation compounds are able to create the rich and unique essential properties, such as, optical absorption spectra, single- & many-particle Coulomb excitations, and electrical conductivities. Apparently, the Fermi level at the conduction Dirac-cone is responsible for the optical threshold frequency of $2E_F$ under the vertical transitions of valence-and conduction bands. Specifically, the many-body excitonic effects might come to exist under the suitable conditions, as revealed in certain optical measurements. They would reduce the optical gap and enhance the initial absorption structure. It is well known that conduction charge carriers have the outstanding Coulomb response under the dynamic and static perturbations of electron-electron interactions. As a result of the conduction and valence electrons [the π* and π carriers], there exist the diverse [momentum. Frequency]-phase diagrams. That is, certain intraband & interband electron-hole regions and different plasmon modes might appear simultaneously, being sensitive to the Fermi level/carrier density and the band structure. Moreover, the Fermi surface, which is formed by the various Fermi-momentum conduction states, is predicted to generate the Friedel charge density oscillation at the long distance. Apparently, it has a strong effect on transport property, e.g., the residual resistivity related to the electron-impurity elastic scatterings.

The graphite-related materials quite differ from the 3D ternary lithium oxide systems in terms of the $Li^+$-based battery anode. For there are certain important differences between the layered $LiC_6/Li^+C_6$ and $Li_8Si_8O_{12}$ bulk compounds. The former are metals/semimetals, while the latter belongs to an indirect-gap insulation with $E_g^i$ more than 5 eV. The interlayer spacings of graphitic sheets, which are determined by the van der Waals interactions of C-$2p_z$ orbitals, are available for the $Li^+$-ion/Li-atom intercalations & deintercalations during the charging & discharging processes. May be, the heterojunction of graphite and electrolyte becomes an issue in greatly enhancing the performance of ion transport. This problem is under the current investigations. But for the latter, there are a lot of chemical bonds within a unit cell, in which the Li- and Si-O bond lengths exhibit the large modulation over 20%. It is deduced to be very easy to change the internal geometric structures [or create the intermediate configurations]. The dramatic transformation between the quasi-stable and intermediate geometries is worthy of the further systematic studies. Of course, the boundary of two distinct lithium oxides needs to be examined thoroughly. Additionally, the semi-metallic, metallic and semiconducting properties, being associated with electrons, might be not play critical roles in ion-transport batteries.

In addition to the lithium-ion-based batteries, there exist the aluminum-[62] and iron-[63] ion-induced ones, especially for the former. The AB-stacked graphite in the second system could serve as the efficient cathode material under the anion transport. Very interesting, the co-existence of $AlCl_4^-$ and $Al_2Cl_7^-$ anions[64] will determine the optimal geometric symmetries between two neighboring graphitic layers and thus the low-lying energy valence and conduction bands. How to mix together for two kinds of large molecular ions could be thoroughly explored by the first-principles calculations, e.g., the diversified distribution configurations associated with the relative ratio. As to the cathode and anode systems of graphite intercalation compounds, their fundamental properties are expected to be totally different from each other in terms of the geometric, electronic, magnetic, optical and transport properties. The intercalations and de-intercalations, which are driven by the aluminum- and chloride-related anions, should be one of the non-negligible transport mechanism in the cathode material of graphite. For example, such behaviors would be responsible for initiating the ionic currents and dominate the stationary ion flows.

## 5. Conclusions

The fundamental properties of (Li,Na,K,Rb,Cs)-intercalated graphite under various stages1-4 have been investigated by mean of first-principle calculations. The weak, but significant Van der Walls interactions would contribute to low-lying π-electronic structure and thus dominate the essential physical properties. The dramatic changes cover the blue shift of the Fermi level/the red shift of the $\sigma$ bands [the n-type doing behaviors], the greatly enhanced asymmetric electron and hole spectra, the obviously reduced conduction electron density for the dilute intercalant systems, the energy spacing between valence and conduction Dirac cones, the initial K/Γ valleys for the π-electronic state with the single/double degeneracy, the whole π band along the K-Γ–K-M- Γ/Γ–K-M-Γ paths [or the H-A-H-L-A/A-L-H-A paths], and a pair of $\sigma$ subbands due to carbon-[2px, 2py] orbitals at the Γ valley of E~-4.0 eV.

The stable layered structures and the metallic/semi-metallic properties are responsible for the important differences with the Li$^+$-based battery anode of the insulating bulk Li$_4$Ti$_5$O$_{12}$. The alkali charge transfers to the carbon sp$^2$ honeycomb lattices is deduced to be large, being comparable with that in 1D graphene nanoribbons[65]. That is to say, the free electron density is roughly identical to the alkali-atom concentration. Conduction electrons have been predicted to have strongly affect the other essential properties, e.g., the great enhancement of the optical threshold frequency, and the various [momentum, frequency]-phase diagrams of Coulomb excitations with the rich single- and many-particle modes.

Most important, the tight-binding model, which could be built from the low-lying VASP band structures under the simultaneous considerations of intralayer C-C & A-A bondings and interlayer C-C and A-C interactions, would be very useful in the diversified phenomena, e.g., the rich magnetic quantization in layered systems only available by this method.

**Acknowledgements:** This work is supported by the Hi-GEM Research Center and the Taiwan Ministry of Science and Technology under grant number MOST 108-2212-M-006-022-MY3 and MOST 108-3017-F-006-003.

# Table

Table 7.1: The stage-dependent ground state energies and the optimal geometric parameters of the stage-n graphite alkali-intercalation compounds.

| | Stage-1 , Alkali-doped Griphite | | | | |
| --- | --- | --- | --- | --- | --- |
| | LiC6 | NaC8 | KC8 | RbC8 | CsC8 |
| distance between alkali metal and carbon layer (Å) | 1.9075 | 2.3005 | 2.662 | 2.9127 | 3.016 |
| Ground State Energy(eV) | -57.29 | -74.96 | -75.13 | -75.04 | -75.12 |

| | Stage-2 , Alkali-doped Griphite | | | | |
| --- | --- | --- | --- | --- | --- |
| | LiC12 | NaC16 | KC16 | RbC16 | CsC16 |
| distance between alkali metal and carbon layer (Å) | 1.9685 | 2.4833 | 2.8233 | 2.9685 | 3.0351 |
| Ground State Energy(eV) | -112.52 | -148.78 | -148.93 | -148.84 | -148.94 |

| | Stage-3 , Alkali-doped Griphite | | | | |
| --- | --- | --- | --- | --- | --- |
| | LiC18 | NaC24 | KC24 | RbC24 | CsC24 |
| distance between alkali metal and carbon layer (Å) | 2.783/1.289 | 2.4295/1.9934 | 2.6766/1.7498 | 2.8382/1.8016 | 2.5999/2.0571 |
| Ground State Energy(eV) | -166.99 | -222.64 | -222.81 | -222.69 | -222.79 |

# Figure Captions

Figure 7.1: The stage-n graphite alkali-intercalation compounds: (a) n=1, (b) n=2, (c) n=3, and (d) n=4.

Figure 7.2: The planar structures of (a) $LiC_6$ and (b) $XC_8$ (X=Na, K, Rb and Cs).

Figure 7.3: The rich band structures of the pristine graphite and stage-1 graphite alkali-intercalation compounds (a) within the corresponding first Brillouin zones: (b) AA & (c) AB stackings without intercalations, (d) $LiC_6$, (e) $NaC_8$, (f) $KC_8$, (g) $RbC_8$ and (h) $CsC_8$.

Figure 7.4: The unusual electronic energy spectra of the AB-stacked stage-2 graphite alkali-intercalation compounds: (a) the reduced first Brillouin zones of AB stacking, (b) $LiC_{12}$, (c) $NaC_{16}$, (d) $KC_{16}$, (e) $RbC_{16}$, and (f) $CsC_{16}$.

Figure 7.5: The rich band structures for the AB-stacked stage-3 graphite alkali-intercalation compounds: (a) the reduced first Brillouin zone, (b) $LiC_{18}$, (c) $NaC_{24}$, (d) $KC_{24}$, (e) $RbC_{24}$, and (f) $CsC_{24}$.

Figure 7.6: The atom- and orbital-decomposed density of states for the (a) AA- & (b) AB-stacked graphite systems, (c) $LiC_6$, (d) $NaC_8$, (e) $KC_8$, (f) $RbC_8$, (g) $CsC_8$, (h) $LiC_{12}$, (i) $NaC_{16}$, (j) $KC_{16}$, (k) $RbC_{16}$, and (l) $CsC_{16}$. The black, red, light blue, pink, gtreen, deep blue, and purple curves, respectively, C-2s, C-2px, C-2py, C-2pz, alkali, carbon, and compound.

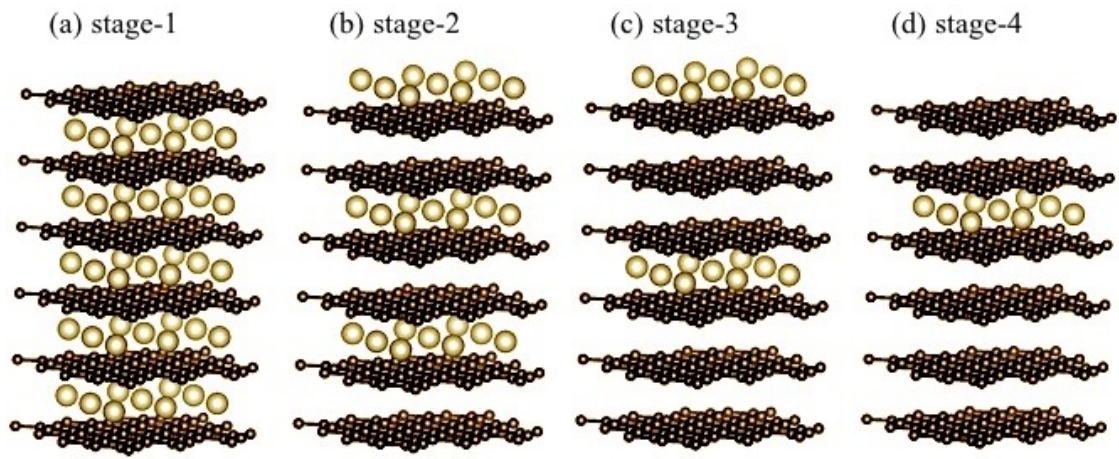

Figure 7.1: The stage-n graphite alkali-intercalation compounds: (a) n=1, (b) n=2, (c) n=3, and (d) n=4.

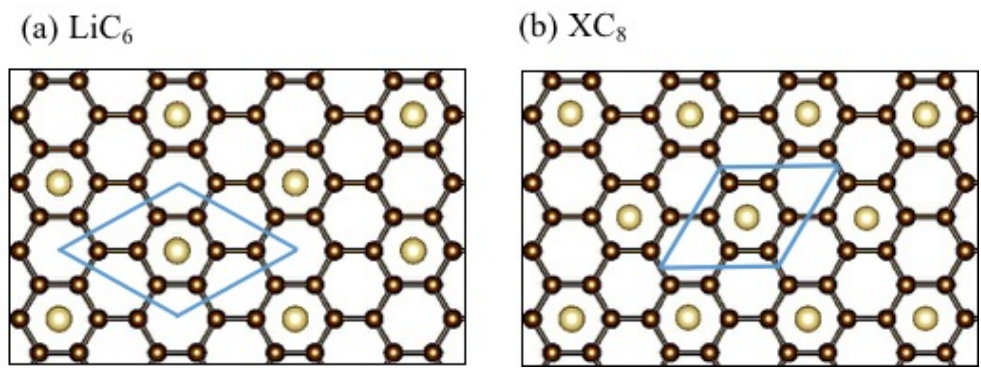

Figure 7.2: The planar structures of (a) $LiC_6$ and (b) $XC_8$ (X=Na, K, Rb and Cs).

Figure 7.3: The rich band structures of the pristine graphite and stage-1 graphite alkali-intercalation compounds (a) within the corresponding first Brillouin zones: (b) AA & (c) AB stackings without intercalations, (d) LiC$_6$, (e) NaC$_8$, (f) KC$_8$, (g) RbC$_8$ and (h) CsC$_8$.

Figure 7.4: The unusual electronic energy spectra of the AB-stacked stage-2 graphite alkali-intercalation compounds: (a) the reduced first Brillouin zones of AB stacking, (b) $LiC_{12}$, (c) $NaC_{16}$, (d) $KC_{16}$, (e) $RbC_{16}$, and (f) $CsC_{16}$.

Figure 7.5: The rich band structures for the AB-stacked stage-3 graphite alkali-intercalation compounds: (a) the reduced first Brillouin zone, (b) LiC$_{18}$, (c) NaC$_{24}$, (d) KC$_{24}$, (e) RbC$_{24}$, and (f) CsC$_{24}$.

Figure 7.6: The atom- and orbital-decomposed density of states for the (a) AA- & (b) AB-stacked graphite systems, (c) $LiC_6$, (d) $NaC_8$, (e) $KC_8$, (f) $RbC_8$, (g) $CsC_8$, (h) $LiC_{12}$, (i) $NaC_{16}$, (j) $KC_{16}$, (k) $RbC_{16}$, and (l) $CsC_{16}$. The black, red, light blue, pink, gtreen, deep blue, and purple curves, respectively, C-2s, C-2px, C-2py, C-2pz, alkali, carbon, and compound.